\documentclass[12pt,preprint]{aastex}

\begin{document}

\title{Does energy of type IIP supernovae depends on stellar mass? }

\author{
N. N. Chugai\altaffilmark{1},
and
V. P. Utrobin\altaffilmark{2}
}
\altaffiltext{1}{Institute of Astronomy, RAS, Pyatnitskaya 48, 109017 Moscow, Russia;
   nchugai@inasan.ru}
\altaffiltext{2}{Institute of Theoretical and Experimental Physics,
   117218 Moscow, Russia; utrobin@itep.ru}

\begin{abstract}

We determine the oxygen density in the central zone of nine type IIP 
supernovae (SN~IIP) at the nebular stage using oxygen doublet 
[O\,I] 6300, 6364 \AA. Combined with two available estimates 
these data indicate that oxygen densities on day 300 are 
distributed in rather narrow range $(2.3\pm1)\times10^9$ cm$^{-3}$.
The result does not depend on the distance, extinction, or 
model assumptions. We demonstrate that the found density distribution 
suggests that the explosion energy of SN~IIP increases with the stellar 
mass. 

\end{abstract}

\section{Introduction}

Type IIP supernovae are caused by an explosion related with the 
core collapse of massive stars.
The theory of stellar evolution predicts that SN~IIP progenitors 
reside in the mass ($M$) range of $9-25~M_{\odot}$ \citep{heg03};
the bounds may be in error by 20\%. The ejected mass ($M_e$) is lower 
by the neutron star and the mass lost by the stellar wind.
How do SN~IIP explode --- the question that still remains unresolved.
Two major mechanism are discussed in this  context: neutrino deposition 
\citep{cow66} and magnetorotational explosion \citep{bik71}.
A third interesting scenario suggests the rotational fragmentation of the 
protoneutron star followed by an explosion 
of neutron mini-star of $0.1~M_{\odot}$ \citep{ims92}

Of particular interest for the observational verification of the explosion 
mechanisms could be the relation between the explosion energy and 
progenitor mass.
Common sense arguments suggests that the explosion energy should increase with 
the star binding energy. The latter increases with the 
progenitor mass \citep{woo02}, so one expects in this case that the 
explosion energy (i.e., kinetic energy at infinity) should rise 
with a progenitor mass.
However, the analyses of the neutrino mechanism in the framework of 
the 2D-hydrodynamics with 
a simplified neutrino transfer implies that the explosion energy, on the 
contrary, should decrease with the rising stellar mass at least 
in the progenitor mass range of $15-25~M_{\odot}$ \citep{fry99}.
Recent numerical experiments using one-dimensional hydrodynamics 
with analytical description of the neutrino luminosity predict the
non-monotonic $E(M)$ relation in the mass range of $10-28~M_{\odot}$ 
with the energy variation in the range of $(0.5-2)\times10^{51}$ erg 
\citep{ugl12}.
As to the magnetorotational mechanism, \citet{moi12}
show that the explosion energy monotonically increases with the  core 
mass, if one admits the constant ratio of the rotational-to-gravitational 
energy.

A phenomenological relation between the energy and mass can be infered  
directly from the hydrodynamic modelling of a large sample of SN~IIP.
This sort of the study for the eight SN~IIP indicates a correlation between 
the energy and progenitor mass \citep{utr13}.
The problem is that the explored sample comprises only progenitors 
with masses $>15~M_{\odot}$; whether this stems from the observational 
selection or the mass overestimation, remains an open issue.
A different conclusion is made by \citet{nad03} using 
estimates based on analytical relations between the observables and the 
supernova parameters: a sample of 14  SN~IIP does not show any correlation 
between the explosion energy and the mass (but see \citet{ham03}). 
At present therefore the issue, whether the energy of SN~IIP depends 
on the progenitor mass, remains unclear from both observational and theoretical
point of view.

In the present paper we study the issue of the energy-mass relation 
for SN~IIP using model independent arguments.
For the freely expanding supernova envelope the density in the 
central zone depends on the ejecta mass and energy as 
$\rho\propto M_e/(vt)^3\propto t^{-3}M_e^{5/2}E^{-3/2}$. 
This relation $\rho(E,M_e)$ 
implies that the presence or absence of the correlation between 
$E$ and $M_e$ could be checked by means of the density measurement at 
some fixed stage.

There exists a simple and efficient method for the 
density determination in the SN~IIP enevelope at the nebular stage 
using [O\,I] 6300, 6364 \AA\ doublet.
The red-to-blue flux ratio in the optically thin case is R/B=1/3.
However, in the inner zone of SN~IIP, where most of the  
synthesised oxygen resides, the optical depth in the [O\,I] 6300 \AA\
line may be large. In this case the doublet ratio 
R/B can be larger than 1/3. This effect was observed originally in 
SN~1987A and was used to estimate the density and filling 
factor of oxygen in this supernova \citep{spy91,chu88}.

It is noteworthy, that the doublet ratio value 
can be hampered by the Thomson scattering which gives rise to
the red wing of [O\,I] 6300 \AA\ line and thus the increase of 
R/B ratio at the early nebular stage ($t<200$ d) up to R/B$>1$ 
\citep{chu92}. To determine the oxygen density therefore
one should use late time nebular spectra. On the other hand, 
at the very late epoch the doublet ratio converges to the limit R/B=1/3, 
in which case the optical depth in the 6300 \AA\ line is impossible to recover.
The favorable conditions for the oxygen density determination 
take place at age of 250--400 days.
Surprisingly, despite the significance of the density diagnostics 
upon the bases of the [O\,I] 6300, 6364 \AA\ doublet was recognized 
long ago, until now apart from SN~1987A this method has been applied 
only for SN~1988A and SN~1988H \citep{spy91}.

In this paper we wish to measure oxygen density for a sample of 
SN~IIP with the nebular spectra of good quality  using [O\,I] 6300, 6364 \AA.
As a result, we hope to recover distribution function of oxygen density 
$p(<n)$ for this category of supernovae. The analysis of this 
distribution in terms of a supernova energy and mass hopefully 
will permit us 
to draw the conclusion on the energy-mass relation and to answer the 
question posed by the paper title. We start with the  
conditions in the oxygen line-emitting zone and the method of 
oxygen density measurement. We find then the oxygen density for a 
sample of SN~IIP and finaly present results for the analysis of the 
density distribution function.

\section{Oxygen density measurement}

The line-emitting oxygen in SN~IIP undoubtedly is a product of thermonuclear 
helium burning in the progenitor core. Prior the supernova explosion this 
oxygen forms the oxygen shell on top of the iron core with 
the $\sim0.2~M_{\odot}$ of silicon in between \citep{woo02}.
Pre-SN models predict that the oxygen mass fraction in the 
oxygen material is $\approx0.8$ \citep{eas94}. The additional 
constituents are Si, Mg, Ne, and C, with Si being dominant in the inner 
and C in the outer layers. The explosion brings about the macroscopic 
mixing of the oxygen material with the outer helium and hydrogen matter 
mediated by Rayleigh-Taylor instability \citep{mue91}.
As a result, the oxygen turns out to be distributed in 
condensations imbedded in the helium and hydrogen material.
Similar oxygen condensations are seen, e.g., in the Cas A supernova remnant 
although with higher velocities compared to SN~IIP because 
the Cas A supernova was SN~IIb \citep{kra08} lacking hydrogen envelope.
The [O\,I] doublet in SN~1987A indicates that the filling factor 
of oxygen condensations is $f\approx0.1$ \citep{spy91,chu88}.
Similar estimate is obtained 
from the flux fluctuations in the doublet profile \citep{chu94}.

The [O\,I] 6300, 6364~\AA\ forbidden lines are emitted by 
transitions between the upper $^1$D$_2$ level and two lower levels,  
$^3$P$_2$ (ground level) and $^3$P$_1$. Generally,  
comparable rates of radiation and collisional transititions suggest that  
the population of three lower levels $^3$P$_{0,1,2}$ should be close 
to the Boltzmann one. However, given the significance of 
our diagnostics one needs to be sure that the assumption of 
the Boltzmann distribution for lower levels is justified. 

We first consider the ionization and thermal balance.
The ionization and heating of oxygen at the nebular stage is maintained 
by gamma-rays from the radioactive decay $^{56}$Co --- $^{56}$Fe. 
The fraction of the deposited energy of gamma-rays spent on the heating 
($\eta$) is taken according to \citet{kfr92}, while the energy 
spent on the ionization and excitation is assumed to be similar, 
i.e., $0.5(1-\eta)$. 
The oxygen recombination coefficient is taken according to \citet{shu82}. 
The model parameters are the expansion 
velocity at the outer boundary of the oxygen line-emitting zone ($v$), 
the amount of $^{56}$Ni ($M_{\rm ni}$), and the density. 
Here we adopt the density to be equal to the oxygen density.
The $^{56}$Ni is assumed to be distributed in 
the oxygen zone. The major coolant are the oxygen emission and 
Mg\,II~2800~\AA\ doublet emission; the later contributes less 
than 10\%. Note that the component of the oxygen material 
with the carbon being dominant among minor species, cools efficiently 
by CO emission and does not contribute to [O\,I] doublet emission 
\citep{liu95}. The observed oxygen doublet emission originates  
primarily from the oxygen with Ne, Mg, and Si being dominant  
admixtures where Mg\,II~2800~\AA\ is the strongest emission next to 
the oxygen doublet. 

Our model takes into account a specific property of the 
central zone of SN~IIP which is revealed through the failure 
of the model with a homogeneous mixing of $^{56}$Ni 
to reproduce the initial ($t<300$ d) 
rise of the [O\,I] flux in SN~1987A \citep[e.g.][]{kfr98}.
To account this  rise one needs to admit that the $^{56}$Ni 
is separated from the oxygen clumps by the absorbing layer devoid of 
the oxygen \citep{chu05}. The average optical depth for gamma-quanta 
of this absorbing layer $\tau_{\gamma}$ we express as some fraction 
$\psi$ of the total optical depth of the oxygen-emitting 
zone $\tau_{\gamma}=\psi\tau_{\gamma,0}$. The $\psi$ value can be 
found by comparing a model and observed luminosity of [O\,I] doublet.

The observed luminosity of the [O\,I] doublet in case of SN~1987A on day 
300 is $2\times10^{39}$ erg s$^{-1}$ \citep{men91,han91}.
Adopting $n=2\times10^9$ cm$^{-3}$ on day 300 \citep{spy91},
$M_{\rm ni}=0.075~M_{\odot}$ \citep{sun91,whi91} 
and $v=1750$ km s$^{-1}$ \citep{chu92} we find that the observed 
luminosity of the [O\,I] doublet implies $\psi=0.27$.
In that case the temperature is $T=4110$~K and ionization $x=0.018$.
Below we use $\psi=0.27$ for other SN~IIP as well.
In the case of SN~2004et, the high luminosity SN~IIP \citep{sah06},
adopting $M_{\rm ni}=0.07~M_{\odot}$ and $v=1900$ cm s$^{-1}$ \citep{utr09}
and the oxygen number density concentration 
$n=2\times10^9$ cm$^{-3}$ on day 300 we find 
$T=3880$~K and $x=0.016$. In the oposite case of the low luminosity 
SN~1997D \citep{tur98} with the parameters 
$M_{\rm ni}=0.005~M_{\odot}$ and $v=600$ cm s$^{-1}$ one gets  
$T=4260$~K and $x=0.02$ assuming 
$n=2\times10^9$ cm$^{-3}$ on day 300. Note that the temperature and 
ionization fraction are not sensitive to the density value.
These results for three different SN~IIP show that the oxygen 
at the considered epoch is essentially neutral and the temperature 
is in the range of 3800 -- 4300~K.

Populations of O\,I lower levels are found by solving 
rate equations taking into account four relevant levels 
$^3$P$_{0,1,2}$ and $^1$D$_2$; transition probabilities are from 
Aller (1984). For the temperature range 3800 -- 4300~K and 
ionization 0.01 -- 0.05 the found population ratio $n_2/n_1$ of 
$^3$P$_1$ and $^3$P$_2$ levels obey Boltzmann law with the accuracy 
of 1\%. The upper $^1$D$_2$ level can be underpopulated by 10\% 
compared to Boltzmann; this however does not affect the doublet R/B ratio.

Given small deviation of the $n_2/n_1$ ratio from the Boltzmann distribution 
and weak sensitivity of this ratio to the temperature in the relevant range 
of values we adopt the Boltzmann law and $T=4000$~K for each case below.
With levels $^3$P$_{2}$, $^3$P$_{1}$, and $^1$D$_2$ numbered 
1, 2 and 3, the ratio of intensities ($I_{\lambda}$) becomes 
\begin{equation}
{\rm R/B}= \exp{(E_{12}/kT)}\left(\frac{\lambda_{13}}{\lambda_{23}}
\right)^5
\frac{\left[1-\exp{(-\tau_{23})}\right]}{\left[1-\exp{(-\tau_{13})}\right]}\,,
\label{eq-rtv}
\end{equation}
where $\tau_{13}$ and $\tau_{23}$ are the optical depth of 
6300 and 6364~\AA\ lines respectively. The line optical depth of 
the homologously expanding envelope in the Sobolev approximation is 
determined by the local oxygen number density $\tau\propto nt$.
This means that the equation (\ref{eq-rtv}) permits us to find the 
doublet ratio R/B given the number density and, on the reverse, to determine 
the number density given the R/B ratio.

The measured doublet ratio and the inferred oxygen number density 
for nine SN~IIP are
listed in Table 1 along with the spectra epochs (column 2). Note, the 
number density are recalculated for the epoch 300 d using the relation 
$n\propto t^{-3}$. The doublet components are partially overlapped in all 
but two supernovae (SN~1997D and SN~2009N), so we use the 
decomposition procedure with the template line profile 
$f\propto\exp{(-|x|^a)}$, 
where $x$ is the wavelength displacement relative to the line center 
in units of the line width ($\Delta \lambda/\Delta \lambda_D$); note, 
the line center may be not at the rest wavelength. The power index 
$a$ in most cases is close to two.
In some cases the doublet lines are slightly asymmetric which  is taken into 
account by introducing a second component. We show in Fig. 1 examples of fits for 
SN~1987A and SN~2012A  with one and two line components respectively.

The doublet ratio from Table 1 are plotted in Fig. 2 along with 
the model R/B ratio for several values of number density on day 300.
The data of Table 1 and Fig. 2 show that the densities of eight SN~IIP 
fall into the narrow range $(1-3)\times10^9$ cm$^{-3}$ and only one 
(SN~2005cs) has $n\approx5\times10^9$ cm$^{-3}$.
Remarkably, the number density of oxygen 
$1.9\times10^9$ cm$^{-3}$ for SN~1988A and 
$1.8\times10^9$ cm$^{-3}$ for SN~1988H (both type IIP) recovered 
by \citep{spy91} fall into the above narrow range of oxygen densities.

Differential ($dN/d\log{n}$) and cummulative $p(<n)$ density 
distributions with SN~1988A and SN~1988H 
taken into account are shown in Fig. 3. The average density value is 
$2.3\times10^9$ cm$^{-3}$ with the the standard deviation 
of $10^9$ cm$^{-3}$.
The relatively small scatter of density values for the sample of SN~IIP
of different luminosity, energy, and ejecta mass is 
rather suprising result. Note that the found density 
distribution does not depend on the distance, extinction, or other 
assumptions and should be considered a reliable fact.

\section{Analysis of distribution function}

\subsection{General consideration and toolkit}

The observation that the oxygen density on day 300 for different 
SN~IIP lie in the 
narrow range already indicates monotonically rising function $E(M_e)$. 
Indeed, as mentioned 
the density in the central zone of the envelope depends on the energy 
and mass 
as $\rho\propto M_e^{5/2}E^{-3/2}$. In the absence of any correlation between 
$E$ and $M_e$ the mass variation by a factor of two and energy by a 
factor of ten 
should cause the density variation by two order, in strong disagreement with 
the found density distribution $p(<n)$. The energy-mass relation 
$E\propto M_e^{5/3}\rho^{-2/3}$ that follows from above expression 
combined with 
the small variation of $\rho^{2/3}$ suggests that the energy 
should increase with the mass of the SN~IIP envelope.

To make more detailed conclusions one needs to perform modelling the 
distribution function $p(<n)$ in the framework of reasonable assumptions.
For our purpose the supernova density-velocity distribution can 
be described analytically 
$\rho=\rho_0/(1+x^k)$ where $x=v/v_0$ while $v_0$ and $\rho_0$ are 
determined by 
parameters $E$ and $M_e$. This function reproduces essential features of 
the density distribution in hydrodynamic models of SN~IIP, particularly, 
a plateau in the central zone and a power law density drop in outer layers 
$\rho\propto v^{-k}$ where $k\approx 8$ \citep{utr07}; below we use $k=8$.

One has to take into account that the oxygen density exceeds  
the average density in the line-emitting zone. The reason is that 
oxygen condensations form as a result of collision of the oxygen shell 
accelerated by the shock with the rarefied helium and hydrogen envelopes.
The deceleration of the oxygen shell is accompanied by the Rayleigh-Taylor
instability followed by the emergence of "fingers" of dense oxygen material 
protruding into the He/H matter. According to 2D-simulation of SN~1987A 
explosion \citep{mue91} the density contrast of oxygen fingers relative 
to the ambient material is $\sim 3...~10$. A similar number can be 
obtained from 
the following arguments. The main stage of the Rayleigh-Taylor 
instability 
coincides with the maxumum deceleration of the oxygen shell. For this 
to be the case the forward shock should sweep up the mass exceeding the 
mass of the oxygen shell $M_{sw}=\mu M_{\rm O}$, where presumably
$\mu\sim 2$. 
In case of SN~1987A the filling factor of the oxygen condensation is 
$f\approx 0.1$ \citep{chu88,spy91}, which implies that 
the oxygen density contrast relative to the ambient material 
in the mixing zone is $(1-f)/f\mu\sim5$, consistent with 2D simulations. 
This value corresponds to the contrast relative to the average density 
$\chi=1/f(1+\mu)\sim 3$.

Another estimate of this value can be made via the comparison of the average 
density in the inner zone of a normal SN~IIP, e.g., SN~1999em calculated by 
one-dimensional hydrodynamic model and the oxygen density derived from 
the [O\,I] doublet profile. Parameters of SN~1999em, $M_e=19~M_{\odot}$ and 
$E=1.3\times10^{51}$ erg \citep{utr07}, with above 
analytical density distribution 
result in the average density in the central zone 
$\rho=2.3\times10^{-14}$ g cm$^{-3}$ on day 300. At the same time, [O\,I] 
doublet ratio gives the oxygen density $\rho_c=6.7\times10^{-14}$ g cm$^{-3}$ 
assuming the oxygen abundance of 0.8 (Table 1). This suggests the density contrast 
for the oxygen condensation $\chi=2.9$. Summarizing all estimates we expect  
the density contrast $\chi\sim 3$ for the oxygen condensations in the central 
zone of SN~IIP.

We are interested in the relation between the energy and the progenitor 
mass $M=M_e+M_{ns}+M_w$, where $M_{ns}=1.4~M_{\odot}$ is the neutron star mass 
and $M_w$ is the mass lost through the wind over the stellar life. Evolutionary 
computations by \citep{heg00} with the mass loss rate of \citet{nie90}
can be approximated in the range of 
$10-25~M_{\odot}$ by the relation $M_w=0.4\beta(M/10\,M_{\odot})^3$ with 
$\beta=1$. We however use below $\beta=0.5$. This choice is based on the 
$M_w$ value estimated for SN~2004et with the ejecta of $24~M_{\odot}$ 
\citep{utr09}.

The density distribution $p(<n)$ is modelled by Monte Carlo 
dicing of $E$ and $M$. The initial progenitor masses presumably 
are distributed according 
to the Salpeter law $dN/dM \propto M^{-2.35}$ in the range $M_1<M<M_2$, where 
for the standard case $M_1=9~M_{\odot}$ and $M_2=25~M_{\odot}$. The 
corresponding 
energy range is $(0.2-4)\times10^{51}$ erg according to parameters of 
the SN~IIP
sample studied by the hydrodynamic modelling \citep{utr13}; 
peculiar SN~2009kf with very high explosion energy is not included in the 
present sample. Below we also consider other options in addition to standard 
choice of parameters.

\subsection{Density distribution and $E(M)$ relation}

We present here simulations of the  $p(<n)$ distribution for seven options. 
Parameters of first four models are given in Table 2. In the first case 
(model m1) there is no correlation between $E$ and $M$: for each $M$ the 
energy is homogeneously distributed in the range $(0.2-4)\times10^{51}$ erg.
The model m2 (standard model) suggests the power law relation 
$E=E_1(M/M_1)^q$ with the power law index determined by the 
interval boundaries $q=\ln(E_2/E_1)/\ln(M_2/M_1)$. We admit a scatter
of energy $E(M)$ for a given mass in the range $E/s-sE$ where 
$s=1.1$. This scatter mimics an effect of possible variations of 
mass loss, rotation, chemical composition and magnetic field. 
We find however that result is not sensitive to the $s$ value.
The model m3 differs from m2 by the upper boundary of mass 
and energy, $M_2=20~M_{\odot}$, $E_2=2\times10^{51}$ erg, and by the density
contrast $\chi=3$. The model m3 approximately corresponds to the mass 
distribution of progenitors according to archive images \citep{sma09}.
The model m4 differs from the model m2 by higher mass of lower boundary,
$M_1=15~M_{\odot}$, which corresponds to the mass range of the SN~IIP sample 
studied hydrodynamically \citep{utr13}. We considered also the 
model m2s in which the luminosity selection effect for the model m2 
is added. The latter is 
realised via multiplying the Salpeter mass spectrum by a factor 
that suppresses the contribution of low mass supernovae $M<13~M_{\odot}$, 
presumably underluminous SN~IIP. This factor is represented by 
$0.5[1+\tanh(x)]$, where $x=(M-12)/2$ and $M$ is solar unit.

Keeping in mind the recent study of the energy-mass relation 
in the framework of neutrino mechanism \citep{ugl12} we consider 
models m5 and m6 that correspond to the mass and energy parameters from this 
paper. Specifically, models m5 and m6 take into account that in the range of
$10-28~M_{\odot}$ there are two intervals, $15-16.3~M_{\odot}$ and 
$23-26~M_{\odot}$, in which supernovae do not explode. In the range of
$10-15~M_{\odot}$ the energy increases in the interval
$(0.9-1.7)\times10^{51}$ erg;
in the range of $16-23~M_{\odot}$ the energy is uniformly distributed 
in the interval of $(0.6-1.6)\times10^{51}$ erg, while in the range of 
$26-28~M_{\odot}$ the energy is uniformly distributed 
in the interval of $(0.9-1.7)\times10^{51}$ erg. The discribed algorithm 
reproduces the energy-mass relation by \citep{ugl12} with a reasonable 
accuracy. The only difference between models m5 and m6 is the mass loss:
the model m5 has the mass loss adopted by \citep{ugl12}
(cf. their Fig. 3), whereas the model m6 suggests the lost 
mass assumed in our other models 
(e.g. model m2). The adopted density contrast in models m5 and m6 
is $\chi=3.5$.

The calculated oxygen density distributions are shown in Fig. 4 along with 
the observed distribution $p(<n)$. The assumption of the absence of any 
correlation between $E$ and $M$ (model m1) apparently contradicts to 
observations. On the other hand, models m2 and m3 with the power law $E(M)$ 
relation and the density contrast $\chi=3$ and 3.5, respectively, 
agree satisfactory 
with the observed density distribution. We performed simulations
for models m2 and m3 with the $\chi$ value randomly distributed 
in the range $3.5\pm0.5$ 
and $3\pm0.5$ respectively and did not find a noticeable difference with 
the case of fixed contrast values. Note that multiplication of the 
contrast value by some factor, $\phi\chi$, results in the shift of the 
model distribution function by $\log\,\phi $ along the $\log\,n$ axis.
In both m2 and m3 models the energy depends on mass as 
$E\approx2\times10^{50}(M/9\,M_{\odot})^{2.9}$ erg. This suggests that the 
density in the central zone depends on mass as $\rho\propto M^{-1.85}$, 
i.e., the density is larger in low mass SN~IIP. This is consistent with 
the fact that largest density in our smaple shows SN~2005cs with the 
progenitor mass of $9-10~M_{\odot}$ \citep{mau05,liv06}.
The model m2s with the selection effect which reduces the contribution of low 
mass supernovae, $M<13~M_{\odot}$ shifts the model m2 towards the lower density 
by 0.13 dex, i.e., by a factor of 1.35. If an observed distribution 
$p(<n)$ is 
affected by the similar selection, the original distribution can be recovered 
by the shift of the observed distribution towards larger density by 0.13 dex.
The model m4 with mass and energy ranges of the hydrodynamical models for 
the sample of seven SN~IIP shows wide range of oxygen densities incompartible 
with the observed distribution. Density distributions in the models 
m5 and m6 based on the energy-mass dependence from \citep{ugl12}
differ significantly from the observed distribution especially in the 
range of small densities.

We now compare the energy-mass relation predicted by the distribution 
$p(<n)$ and parameters of hydrodynamic modelling. The relation $E(M)$ in the 
standard model m2 is shown on the plot $\log E - \log M$ along with 
parameters of hydrodynamic models \citep{utr13}. We included 
also our parameters of SN~2012A ($M=15.1~M_{\odot}$, $E=5.2\times10^{50}$ erg). 
The progenitor mass is derived from ejecta mass 
$M_e$ combined with neutron star and lost mass calculated 
according expression given in Sec. 3.1. The power law dependence $E\propto M^{2.9}$ 
is qualitatively consistent with the best fit relation 
$E\propto M^{3.8}$ for the hydrodynamic parameters of SN~IIP sample. 
Yet the difference between values of $E$ and $M$ of two sorts of relations 
is apparent. If the difference is entirely due to the mass, then the 
mass values of the SN sample turns out to be shifted towards higher masses
by a factor of $\approx1.3$.

\section{Discussion and conclusions}

The goal of the paper was to determine the oxygen density for a sample 
of SN~IIP using the [O\,I] doublet in nebular spectra and then to  
study the relation between energy and progenitor mass upon the bases 
of the recovered density distribution. It was found unexpectedly  
that the range of oxygen number density on day 300 is very   
narrow $(2.3\pm1)\times10^9$ cm$^{-3}$. Remarkably, this result 
does not depend on distance, extinction, or any assumptions.
The modelling for the found density distribution implies that 
the explosion energy of SN~IIP increases with the progenitor mass. This 
result reflects the important property of an explosion mechanism of SN~IIP  
that should be used to constrain explosion models.
It should be emphasised that we do not include into the family of 
SN~IIP events similar to SN~1994W \citep{sol98,chu04} 
and SN~2009kn \citep{kan12} which mimic SN~IIP by their light curves 
but essentially differ by their spectra.

The power index of the $E(M)$ relation found from the distribution  
$p(<n)$ is close to the tangent of the scattering 
plot of hydrodynamic parameters of SN~IIP on the $\log E-\log M$ plane.
The oxygen density distribution thus confirms the conclusion 
about the increase of the explosion energy with the mass that is 
indicated by parameters of the hydrodynamic models \citep{utr13}.
On the other hand, we find that the $E(M)$ relations obtained from the 
hydrodynamic parameters and from the distribution 
$p(<n)$ differ by values $E$ or/and $M$. Interestingly, in this respect, that 
the observed distribution $p(<n)$ is not consistent with the model in which 
lower boundary is $15~M_{\odot}$, i.e. equals the lower limit of the sample of 
SN~IIP studied hydrodynamically. One of the reason of this inconsistency 
could be that masses derived from hydrodynamic models are overestimated.
Independent argument in favor of this possibility stems from the fact that 
progenitor masses of hydrodynamic models lie in the range $>15~M_{\odot}$, 
notably larger than the lower boundary ($\approx9~M_{\odot}$) which 
theoretically are 
associated with SN~IIP \citep{heg03}. This disparity is strengthend
 by the fact that lower boundary of progenitor masses ($\approx8~M_{\odot}$)
extracted from the archival images \citep{sma09} is close to the 
theoretical lower boundary.

The conclusion that the explosion energy of SN~IIP grows with the 
progenitor mass can serve a good observational test of the explosion theory 
for this category of supernovae. The current state of the theory however 
does not permit us to use this test with all its power. The 
energy-mass relation in the framework of neutrino mechanism recovered  
by \citet{ugl12} using one-dimensional model predicts the 
distribution $p(<n)$ strongly unlike the observed one. Yet existing 
uncertainties of the neutrino mechanism do not rule out the monotonic 
increase of the 
explosion energy in the mass range $10<M<M_{up}$, where $M_{up}$ is 
poorely known value and probably lying in the range of $20...~25~M_{\odot}$ 
(T. Janka, private communication). In the mechanism of the mini-neutron star 
the energy release is invariant, $\approx 10^{51}$ erg 
\citep{ims92}. Since the binding energy grows with the mass this 
mechanism predicts decrease of the explosion energy with the mass and 
therefore cannot be universal for SN~IIP.
Currently, only magneto-rotational explosion admits the energy increase with 
the mass \citep{moi12} which makes it an appropriate mechanism for 
the SNe~IIP.

\vspace{1cm}
We are gratefull to Lina Tomasella for sending us spectra of SN~2012A.

\medskip


\newpage
\clearpage

\begin{figure}
\epsscale{.80}
\plotone{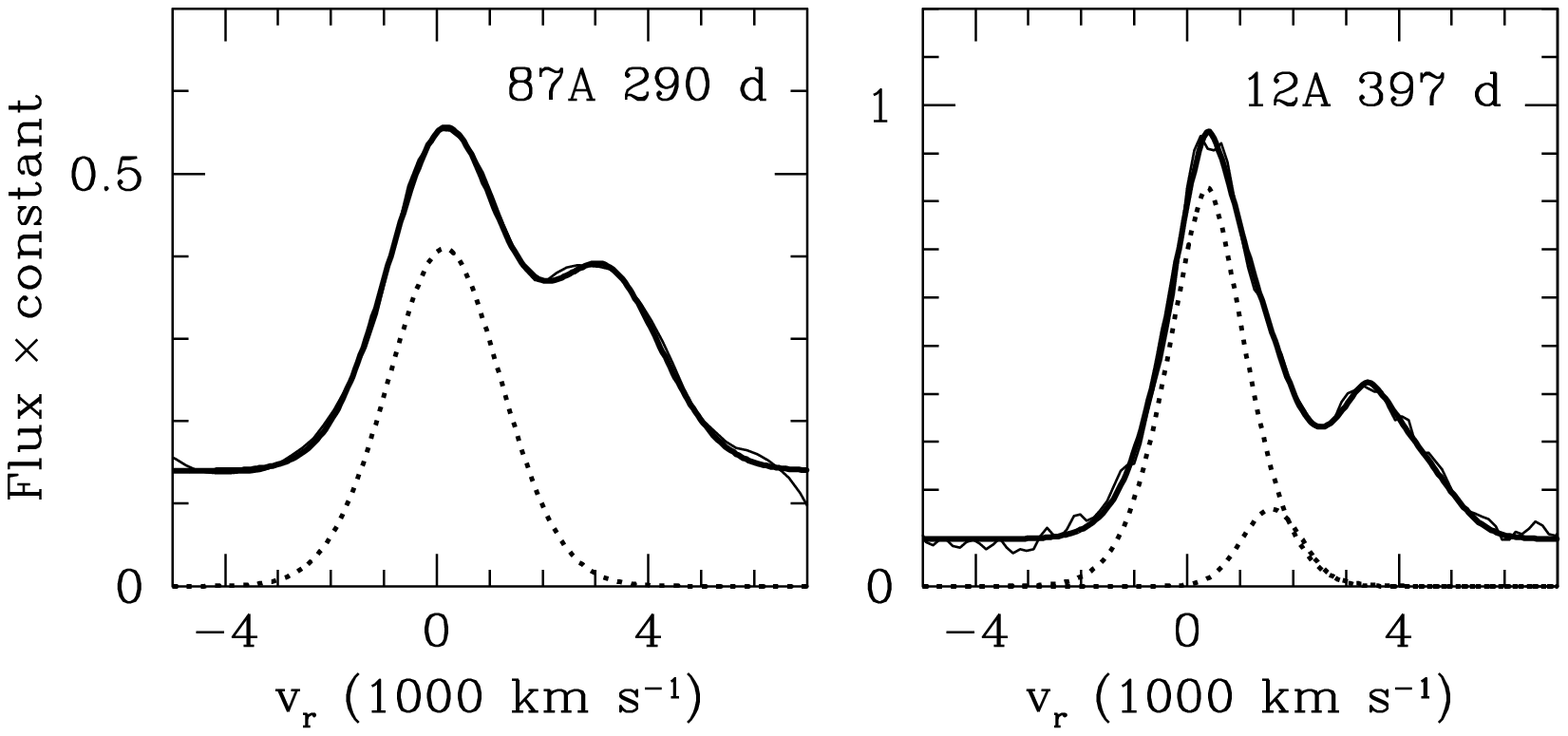}
\caption{Oxygen doublet [O\,I] 6300, 6364 \AA\ in SN 1987A on day 290 and  
SN~2012A on day 397. {\em Thin} line is observations, {\em thick} 
line is the model with the ratio R/V=0.58. {\em Dotted} line shows the 
model of 6300 \AA\ line. In the case of SN~2012A the line is asymmetric 
and this is taken into account by the inclusion of the additional component 
shown by dotted line; For the dominant component R/V=0.38.
	 }
\end{figure}

\clearpage
\begin{figure}
\epsscale{.80}
\plotone{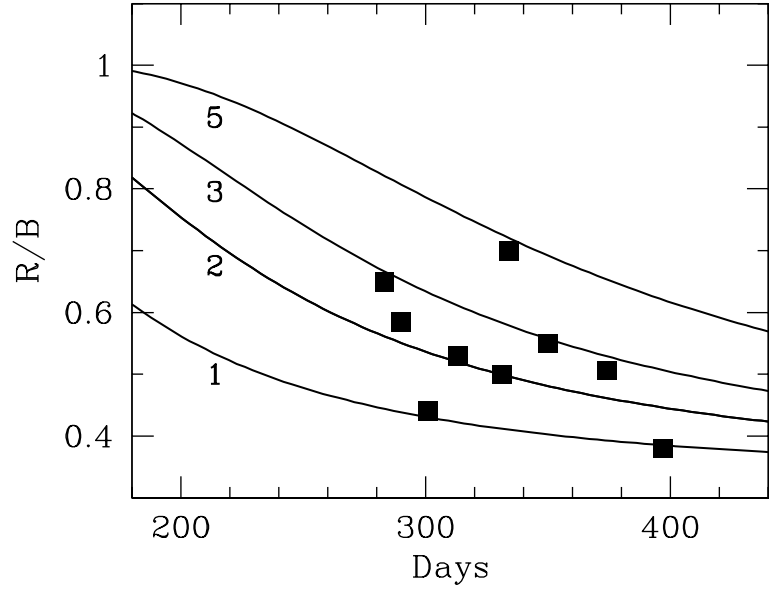}
\caption{Doublet ratio R/V for the SN~IIP sample at the nebular epoch 
(squares). 
Solid lines show the theoretical evolution of the ratio R/B for 
the oxygen number density on day 300 equals 1, 2, 3, and 5 in 
units of $10^9$ cm$^{-3}$. All the supernovae fall into this range, 
while 8 of 9 supernovae fall in the range $(1-3)\times10^9$ cm$^{-3}$.
	 }
\end{figure}

\clearpage

\begin{figure}
\epsscale{.80}
\plotone{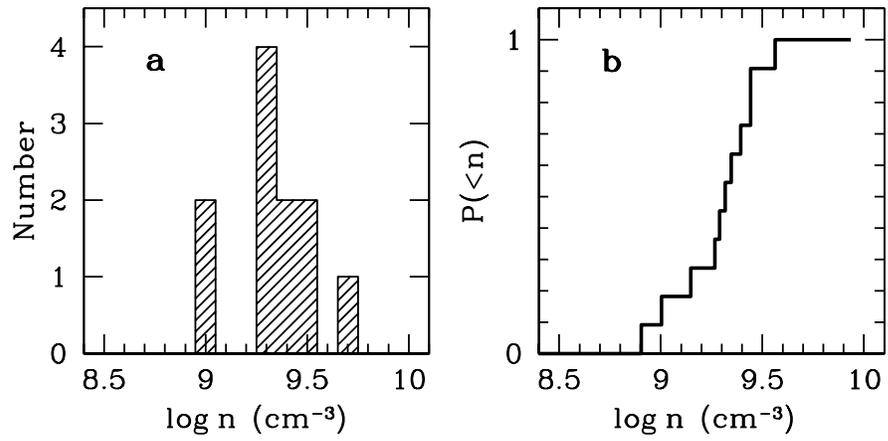}
\caption{Differential ({\bf a}) and cumulative ({\bf b}) distribution of oxygen 
number density for SN~IIP on day 300. The sample from Table 1 is extended by 
the inclusion SN 1988A and SN 1988H (Spyromilio \& Pinto 1991).
	 }
\end{figure}

\clearpage

\begin{figure}
\epsscale{.80}
\plotone{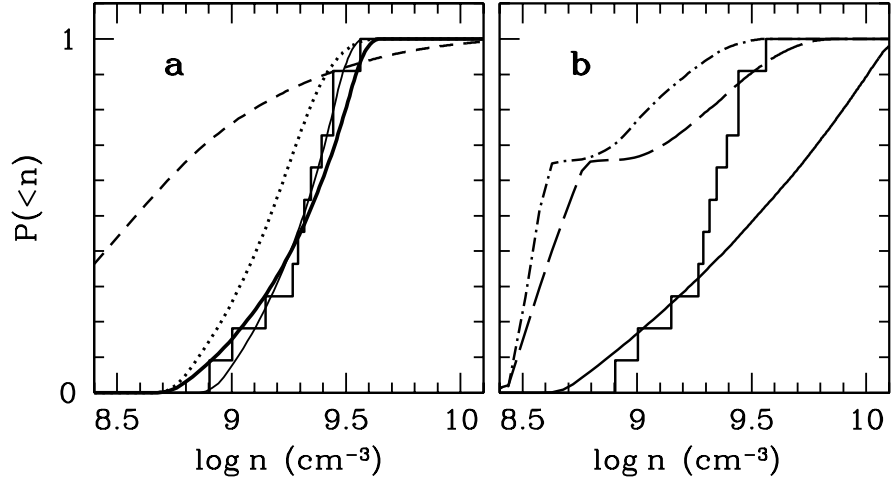}
\caption{Calculated oxygen density distributions of SN~IIP on day 300 compared to 
the observed distribution ({\em histogram}). The panel ({\em a}) shows: 
the model m1 (Table 2) in which no correlation between $E$ and $M$ is assumed 
({\em dashed}) line; the standard model m2 ({\em thick solid} line); 
the model m3 ({\em thin solid} line), and the model m2s with the luminosity 
selection ({\em dotted} line). The panel ({\em b}) shows the model m4 
({\em solid} line), and two other models: m5 with strong mass loss 
({\em dash-dotted} line) and the model m6 with the moderate mass loss 
({\em long dash}) both based on the $E(M)$ relation reported 
by Ugliano et al. (2012).
	 }
\end{figure}

\clearpage

\begin{figure}
\epsscale{.80}
\plotone{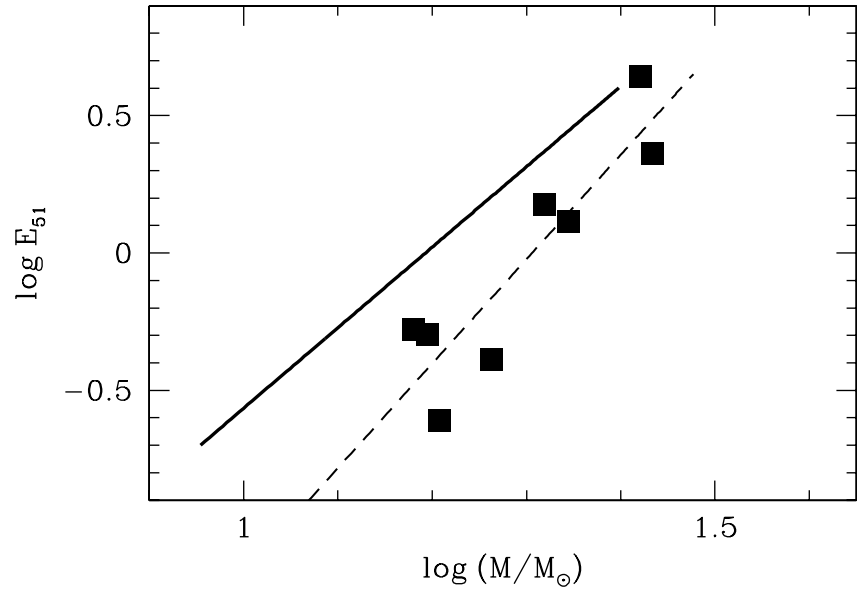}
\caption{Scattering plot $E$ vs. $M$ for eight SN~IIP with parameters measured 
by means of hydrodynamic modelling ({\em squares}). {\em Solid} line is 
the $E(M)$ relation 
of the models m2 and m3 for the distribution $p(<n)$, {\em dashed} line is 
linear fit for the shown SN sample.
	 }
\end{figure}

\clearpage

\newpage
\clearpage
\begin{table}
  \caption{Oxygen number density on day 300 for SN~IIP} 
\medskip   
  \begin{tabular}{ccccl}
\hline

SN    &  $t$     & R/B &     $n(300d)$        & Spectrum \\
      &  day    &     & $10^{9}$ cm$^{-3}$  & \\ 
\hline

 1987A  & 290    & 0.58 &   2.3   &   Phillips et al (1990)\\
 1992H  & 313    & 0.53 &   2.15  &   Clocchiatti et al. (1996)\\
 1997D  & 350    & 0.55 &   2.9   &   Turatto et al. (1998)\\
 1999em & 331    & 0.50 &   2.0   &   Leonard et al. (2002)\\
 2004dj & 287    & 0.65 &   2.9   &   Meikle et al (2011)\\
 2004et & 301    & 0.44 &   1.1   &   Sahu et al. (2006)\\
 2005cs & 334    & 0.70 &   4.6   &   Pastorello et al. (2009)\\
 2009N  & 374    & 0.51 &   2.65  &   Tak\'{a}ts et al. (2013)\\
 2012A  & 397    & 0.38 &   0.92  &   Tomasella et al. (2013)\\
\hline
\end{tabular}
\label{t-mod} 
\end{table} 

\newpage

\begin{table}
  \caption{Parameters of models for oxygen density distribution}
\medskip  
  \begin{tabular}{ccccccc}
\hline

Model & $M_1$ & $M_2$ & $E_1$ & $E_2$  & $\chi$ & Comment\\
  & \multicolumn{2}{c}{$M_{\odot}$} &  \multicolumn{2}{c}{$10^{51}$ erg} &
     &\\ 
\hline

 m1     &   9   & 25  &     0.2    & 4  & 3.5  & $E$ and $M$ do not correlate \\
 m2     &   9   & 25  &     0.2    & 4  & 3.5 & \\
 m3    &    9   & 20  &     0.2    & 2  & 3   & \\
 m4    &    15  & 20  &     0.2    & 4  & 3   & \\
\hline
\end{tabular}
\label{t-mod} 
\end{table} 

\newpage
\clearpage

\end{document}